\documentclass[fleqn,10pt]{wlscirep}

\usepackage{fullpage}
\usepackage{amssymb,amsmath}
\usepackage[utf8]{inputenc}
\usepackage{siunitx}
\usepackage[version=4]{mhchem}
\usepackage{lipsum}
\usepackage{authblk}

\usepackage[left]{lineno}

\usepackage{setspace}
\usepackage{url}

\usepackage{graphicx}

\title{Impact sculpting of the early martian atmosphere}

\author[1,2,*]{Oliver Shorttle}
\author[3]{Homa Saeidfirozeh}
\author[2,4]{Paul Brandon Rimmer}
\author[3, 5]{Vojt\u{e}ch Laitl}
\author[3]{Petr Kubel\'{i}k}
\author[3,6]{Luk\'a\u{s} Petera}
\author[3]{Martin Ferus}
\affil[1]{Institute of Astronomy, University of Cambridge, Madingley Road, Cambridge CB3 0HA, UK}
\affil[2]{Department of Earth Sciences, University of Cambridge, Downing Street, Cambridge CB2 3EQ, UK}
\affil[3]{J. Heyrovsk\'{y} Institute of Physical Chemistry, Czech Academy of Sciences, Dolej\v{s}kova 3, CZ 18223 Prague 8, Czech Republic}
\affil[4]{Cavendish Laboratory, University of Cambridge, JJ Thompson Avenue, Cambridge CB3 0HE, UK}
\affil[5]{University of Antwerp, Faculty of Science, Groenenborgerlaan 171, BE-2020 Antwerpen, Belgium}
\affil[6]{Department of Inorganic Chemistry, Faculty of Science, Charles University, Hlavova 8, Prague, Czech Republic}

\affil[*]{Corresponding author: \url{shorttle@ast.cam.ac.uk}}

\begin{abstract}
{\bf
    Intense bombardment of solar system planets in the immediate aftermath of protoplanetary disk dissipation has played a key role in their atmospheric evolution.  During this epoch, energetic collisions will have removed significant masses of gas from rocky planet atmospheres.  Noble gases are powerful tracers of this early atmospheric history, xenon in particular, which on Mars and Earth shows significant depletions and isotopic fractionations relative to the lighter noble gasses.  To evaluate the effect of impacts on the loss and fractionation of xenon, we measure its ionization and recombination efficiency by laser shock and apply these constraints to model impact-driven atmospheric escape on Mars.  We demonstrate that impact bombardment within the first $200$--$300\,\text{Myr}$ of solar system history generates the observed Xe depletion and isotope fractionation of the modern martian atmosphere.  This process may also explain the Xe depletion recorded in Earth's deep mantle and provides a latest date for the timing of giant planet instability.
}
\end{abstract}
\begin{document}

\maketitle

\section*{Introduction}
Noble gases are a key tracer of the early loss and gain of planetary atmospheres.  Their inert nature makes their abundances and isotopic composition ideal archives of fractionation processes that occur during volatile loss and redistribution in planets.  Xenon, in particular, has proven an enigmatic carrier of information on atmospheric evolution, showing anomalous depletions and mass fractionations in the atmospheres of both Earth and Mars \cite{pepin1991_icarus}.   On Earth, this has led to the identification of the `missing Xe' problem, where Xe is noted to be unusually depleted in Earth's atmosphere given the element's high mass  \cite{ozima2002_cup}.  Associated with this depletion is a large mass-dependent fractionation of Xe isotopes, greater than that seen among the isotopes of Kr despite their being lighter, and, remarkably, is a signal that has been constrained to have grown to its modern value over the first $2\,\rm{Gyr}$ of Earth history \cite{pujol2011_epsl,avice2018_gca,ardoin2022_gpl}.  

\begin{figure}[!ht]
  \includegraphics[width=\textwidth]{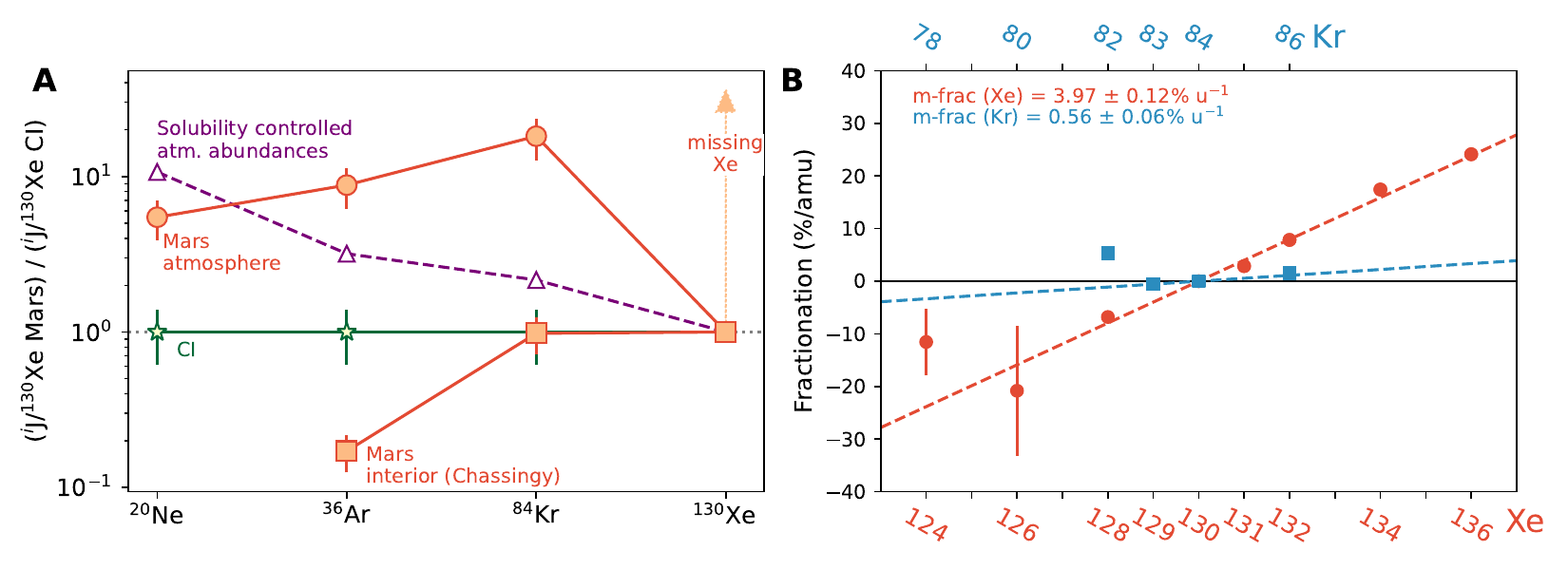}
  \centering
  \caption{{\bf Key characteristics of the martian noble gas reservoirs. (A)}, Elemental abundances of noble gases in the martian interior (as represented by Chassigny \cite{swindle2002_revmin}) and atmosphere \cite{pepin1991_icarus,bogard1998_gca}, ratioed to the abundance of Xe, relative to the elemental ratio in CI chondrites \cite{swindle2002_revmin}.  Xe is noticeably depleted compared to the other noble gases.  The relative noble gas abundances that would be produced in an atmosphere formed from a two stage degassing process are shown as triangles.  First, during the magma ocean stage, noble gases are distributed between the atmosphere and interior according to their relative solubilities \cite{carroll1994_revmin}, after which the initial atmosphere is lost.  Second, subsequent volcanic degassing quantitatively removes noble gases from the melted sources region as magmas degas at low pressure.  In this way, the most soluble gasses are dominant in the atmosphere that is restored.  {\bf (B)}, The isotopic fractionations of Xe (red circles) and Kr (blue squares) in Mars's atmosphere (data from ref. \citenum{conrad2016_epsl}).  Fractionations were calculated per atomic mass unit, considering only the heaviest four isotope ratios of Xe (131/130, 132/130, 134/130, and 136/130) and the 83/84 and 84/84 ratios of Kr.  All reported errors are 1 sigma.}
  \label{fig:abundances}
\end{figure}

Like Earth, Mars's atmosphere evidences preferential Xe loss and isotopic fractionation compared with other noble gases (e.g., ref. \citeonline{conrad2016_epsl}, Fig. \ref{fig:abundances}).  The isotopic fractionation of Xe is of similar magnitude between the two atmospheres \cite{swindle1986_gca}, however, the atmospheres began their evolution from different starting compositions, solar Xe in the case of Mars \cite{pepin2000_ssr} compared with U-Xe for Earth (itself a mixture of cometary and chondritic Xe, \cite{pepin1991_icarus,marty2017_science}).  Importantly, the isotopic fractionation of Xe in the martian atmosphere must have occurred early: data from two martian meteorites, ALH 84001 and NWA 11220, which sample Mars's atmosphere at $\sim4.2\,\rm{Ga}$ and $\sim4.4\,\rm{Ga}$ respectively, indicate that fractionation to the modern value occurred rapidly in the first few hundred million years of the planet's history \cite{cassata2017_epsl,cassata2022_epsl}.

An explanation for the progressive multi-billion year isotopic fractionation of Earth's atmospheric Xe has been provided by ref. \citenum{Zahnle2019_gca}, who suggest that Xe escaped Earth in a photo-ionized hydrogen wind.  In this model, Xe loss and fractionation occur sporadically over the first several billion year's of Earth history and only terminate when atmospheric oxygenation sharply decreases the $\rm{H_2}$ mixing ratio of the atmosphere.  Key to this model is the low ionization potential of Xe, the lowest among the noble gases, which means it is preferentially removed despite its high mass: ionized Xe couples to open magnetic field lines and is dragged to space by ionized hydrogen. A similar model of Xe escaping in an ionized wind has recently been advanced to explain Xe's loss from Mars's atmosphere \cite{cassata2022_epsl}.  In the case of Mars, observations require that the fractionation terminates early (before $4.2\,\rm{Ga}$, \cite{cassata2017_epsl}), which is suggested to occur because of either waning extreme solar UV, diminished $\rm{H_2}$ sources on the planet, or cessation of the martian magnetic field \cite{cassata2022_epsl}. The latter scenario is disfavoured because of evidence of magnetism in rocks younger than 4.2\,Ga \cite{weiss2002_epsl}.  Diminishing $\rm{H_2}$ sources on the planet may be inconsistent with requirements for Mars's transiently wet early climate, which recent modelling has suggested needed episodic reducing conditions (i.e., high-$\rm{H_2}$) for over a billion years into the planet's life \cite{wordsworth2021_ngeo}.  This is long after fractionation of Xe in the atmosphere is observed to have ceased \cite{cassata2017_epsl}.  

Intriguingly, recent work has indicated that a separate epoch of preferential Xe loss must have also occurred much earlier in Earth's history, earlier than the slow loss recorded in its atmosphere over $\sim{}2\,\rm{Gyr}$ \cite{peron2022_epsl}.  Earth's deep mantle may, therefore, have captured evidence of a separate Xe (and wider volatile) loss episode within its first few hundred Myr \cite{mukhopadhyay2012_nature}; an episode contemporaneous with Mars's observed Xe loss \cite{cassata2017_epsl}.  Together, these observations suggest that additional processes(s) may have operated across the solar system early in the life of rocky planets to drive the rapid loss of volatile elements.  As Xe records preferential depletion by these processes, models other than the classic hydrodynamic escape scenarios are required (e.g., \cite{hunten1987_icarus}), as these would favour the preferential loss of light gases over Xe.

Although long considered in the context of atmospheric loss\cite{zahnle1993_jgr}, impact bombardment is one mechanism for isotopic fractionation of atmospheres that has received little attention.  The impact bombardment of planets is widely recognised as an important process shaping their formation and evolution \cite{ahrens1993_annrev}.  During planetary growth, accretionary impacts are the primary mechanism by which rocky planets gain mass \cite{safronov1972_book,greenberg1978_icarus,rubie2015_icarus}, with the most dramatic and recent known example in the solar system being the collision of a Mars-sized impactor with the proto-Earth to form the Moon \cite{cuk2012_science}. One peak in impact rates onto planets likely occurs following the dissipation of their natal protoplanetary disk, which in the solar system occurred at $\sim3\,\rm{Myr}$ after its birth \cite{weiss2021_sciadv}.  Later peaks in impact fluxes would coincide with dynamical reorganisations of the solar system giant planets \cite{gomes2005_nature}.  

The abundance and rate of impacts in this early epoch have been constrained by cratering records on the Moon \cite{neukum2001_chron}, Mars \cite{hartmann2001_chron,neukum2001_chron,bottke2007_icarus}, and by the highly siderophile element (HSE) abundances of the terrestrial and martian mantles \cite{bottke2010_science}.  However, the timing of any increased bombardment following giant planet instability remains uncertain: initially linked to clustered ages of lunar rocks between $\sim3.5$--$4.2\,$Ga \cite{tera1974_epsl,bottke2017_annrev}, it now seems likely that the lunar observations at least can be explained with a gradual tailing-off of accretion \cite{morbidelli2018_icarus} and that any instability occurred earlier \cite{nesvorny2018_annrev,nesvorny2018_nast,avdellidou2024_science}.  Impact bombardment could significantly affect atmospheric retention on rocky planets \cite{melosh1989_nature,schlichting2015_icarus,schlichting2018_ssr}, being capable of both delivering and removing volatile elements depending on impactor size.   In this case, the atmospheric evolution of the terrestrial planets may itself be a record of this early bombardment history.

\begin{figure}[!ht]
  \includegraphics[width=\textwidth]{"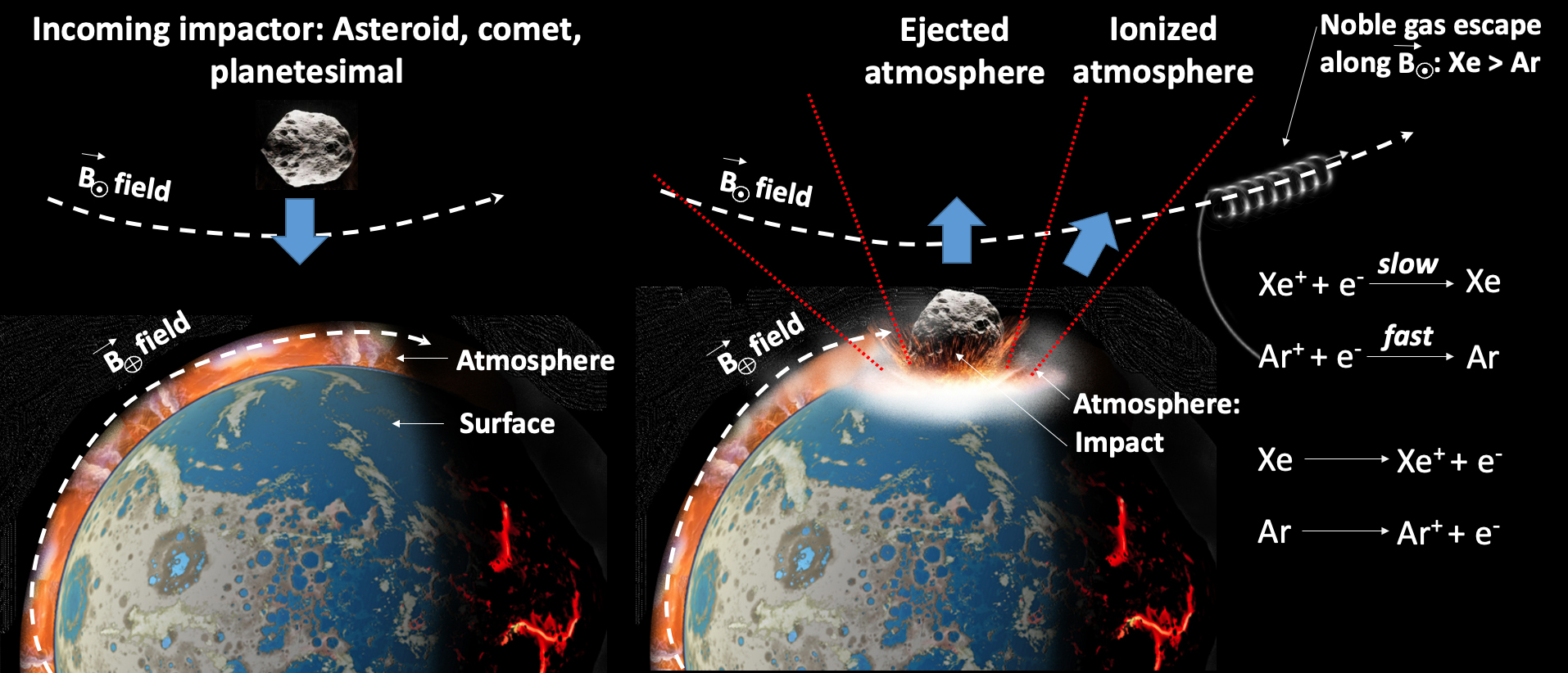"}
  \centering
  \caption{{\bf Model framework for describing Xe-loss and fractionation from the martian atmosphere.} Impacts eject all material in a central cone, with an annulus of partially ionized material escaping the atmosphere according to whether ions survive to high altitude to connect with Mars's open magnetic field lines\cite{connerney2005_pnas}.}
  \label{fig:sketch}
\end{figure}

Impacts have been presumed to not affect the fractionation of Xe in planetary atmospheres because of their propensity to remove atmosphere from their target indiscriminately \cite{walker1986_icarus,zahnle1993_jgr,schlichting2015_icarus}.  However, whilst a hypervelocity impactor is moving through the atmosphere, it produces a high temperature ionized shock (a quasi-neutral plasma over $10^4\,\rm{K}$) in a wide region around it \cite{silber2018_asr}.  This plasma ends up thrown away from the planet's surface to high altitude (Fig. \ref{fig:sketch} and ref. \citeonline{schlichting2015_icarus}).  Whilst for a sufficiently large impactor some atmosphere in this cone-shaped region will be directly ejected, the planet will also retain a proportion of the ionized atmosphere.  The transiently ionized atmosphere from hypervelocity impacts, therefore, represents a potential mechanism of creating ionized Xe at heights in the atmosphere where it may couple to magnetic field lines and escape the planet.  This is a scenario similar to that envisaged by ref. \citenum{Zahnle2019_gca}, but with impacts rather than photons as both the source of the ionization and as the mechanism to move Xe up to high altitude (i.e., induce enhanced vertical mixing).  Importantly, Xe loss by impacts would decouple the history of Xe from the $\rm{H_2}$ mixing ratio of a planetary atmosphere: whereas ref. \citenum{Zahnle2019_gca, cassata2022_epsl} have emphasised the importance of hydrogen for ionising the Xe and transporting it vertically. Impacts achieve both of these processes, ionising Xe through high temperature plasma and ballistically lofting it in the atmosphere, without recourse to specific background atmospheric $\rm{H_2}$ mixing ratios.  This is perhaps useful in the context of the requirements of Mars's early climate \cite{wordsworth2021_ngeo}, and would present a new dependence for the timescale of Xe's evolution, occurring during epochs of intense impact bombardment.  Indeed, such an association between impacts and Xe isotopic evolution has been noticed for Earth (Fig. 7c of \cite{avice2018_gca}), although our present work does not seek to explain Earth's more protracted Xe-loss history.

For such a scenario to explain the relative Xe abundance and isotope composition of Mars's atmosphere, it is key that Xe, compared with the other noble gases, be preferentially affected by impact-driven ionization and escape.  Two conditions would contribute to preferential Xe loss and isotope fractionation: first, if the high temperatures of an impact plasma ionized Xe appreciably more than other noble gases; second, preferential Xe loss would be aided by the recombination rate of Xe being lower than that of the other noble gases, given that the ionized Xe will also need to be transported from the site of the shock to a height where it can couple to open magnetic field lines and be lost. In particular, for this second condition, we are interested in the rate of the reaction
\begin{equation}
\ce{Xe^+} + e^- + \ce{M} \rightarrow \ce{Xe} + \ce{M},
\label{eq:recomb}
\end{equation}
where \ce{M} is any neutral third body, versus the rate of equivalent reactions for other noble gases. If one or both of these conditions were met without the other actively disfavouring Xe ionization, then more Xe remains ionized once lofted up by impact than other noble gases; aiding Xe escape.

Experiment and theory constrain the ionization energy for xenon to be lower than for any other noble gas \cite{NIST_asd} and that the rate at which ionized xenon recombines with its electron is slower than for the other noble gases \cite{anicich1993survey}. However, the experiments that give these rates either measure the thermal ionization rate and calculate the rate of the reverse reaction \cite{Johnston1961,Kornegay1963}, or have only measured the forward electron-ion recombination rates \cite{Vitols1973,Barbet1975}. A final limitation of the current data is that no ion-electron recombination experiments have been performed for noble gases in the context of a planetary atmosphere; the role of geologically relevant background gasses in Xe recombination is therefore poorly constrained.

To provide new constraints on the key parameters of impact-driven ionization, we performed laser shock experiments to simulate impactor entry into planetary atmospheres. These give us direct experimental constraints on the thermal ionization of noble gases during an impact and the recombination rates for those ions during, and immediately after, the impact. We obtain new estimates of ionization fractions and recombination rates for Xe, using Ar as a reference noble gas with a higher ionization threshold.  Combining the Xe ionization parameters with a simple model of impact driven atmospheric loss (based on ref. \citeonline{schlichting2015_icarus}), we then explore the viability of impacts as the exclusive mechanism explaining Mars's history of Xe isotopic fractionation.  We find that impacts can explain the early onset and magnitude of Xe isotopic fractionation in the martian atmosphere, subject to Mars's initial mass of the atmosphere and the specific bombardment history it experiences.  Whilst other loss processes were likely simultaneously operating on planets to drive atmospheric loss and Xe fractionation \cite{cassata2022_epsl}, our results suggest that impact sculpting of the early martian atmosphere was likely significant.  In this context, the recently discovered Xe-depletions in the deep Earth \cite{peron2022_epsl} indicate impact-driven loss processes may have operated early in its history as well.

\section*{Results}

\subsection*{Laser shock experiments}
\begin{figure}[!ht]
  \includegraphics[width=1\textwidth]{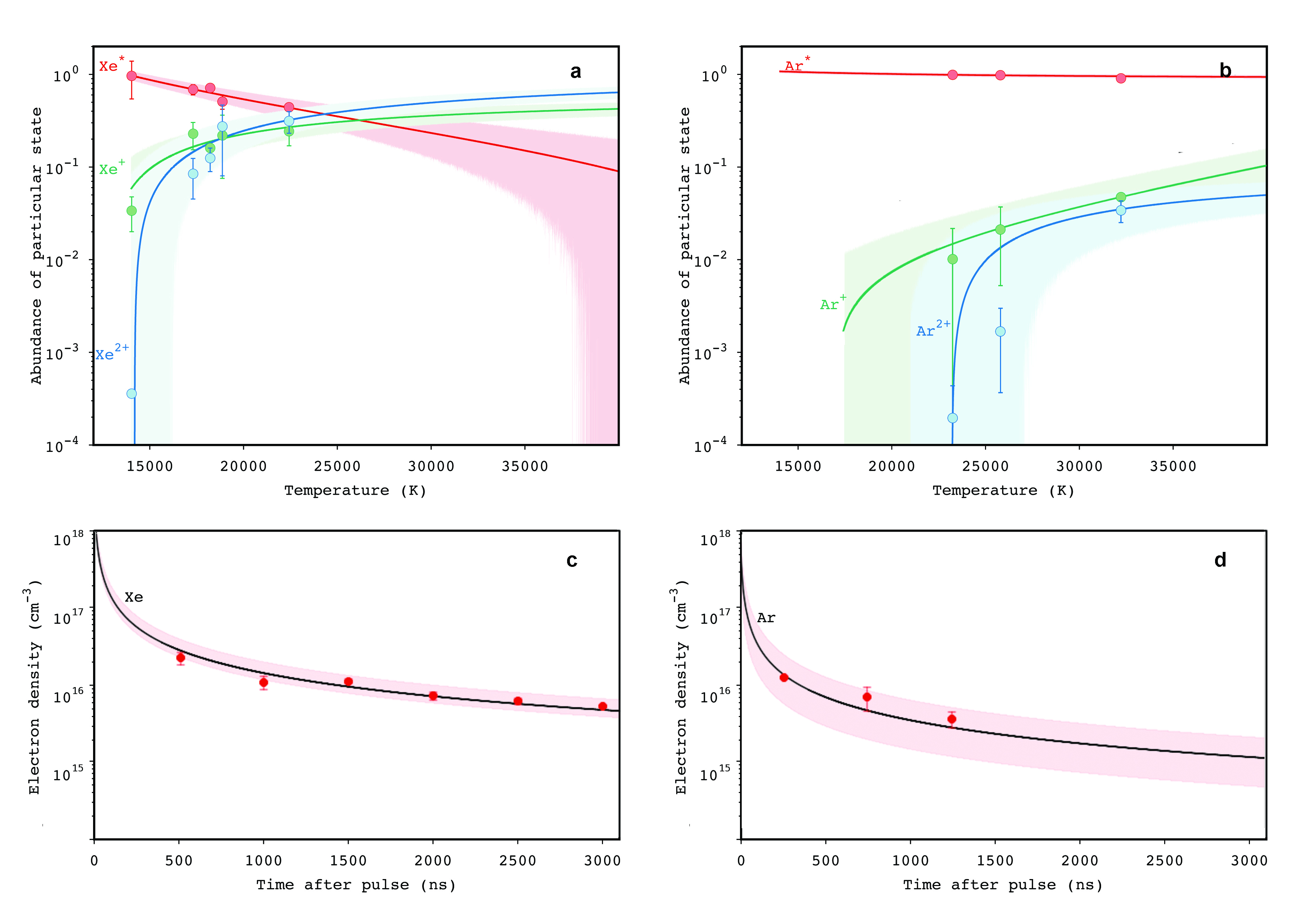}
  \centering
  \caption{{\bf Experimental results showing the relative ionisation efficiency and recombination rates of Xe and Ar during laser shock.} {\bf (A)} and {\bf (B)}, represent the abundance of particular Xe/Ar states as a function of temperature for experiments run with pure Xe and Ar at a total pressure of $\sim0.17\,\text{bar}$ at $1500\,\rm{ns}$ following the laser pulse.  Superscript star indicates neutral species, and superscript plus indicates charged or multiply charged species. The solid lines show the fit from simulations of the excited states, with shaded envelopes giving the 97.5\% confidence interval from data uncertainty: simulation were based on modelling Saha ionisation equilibrium with varying temperature (see Materials and Methods for details). ({\bf C}) and ({\bf D}) Show the electron density variation with time for pure Xe/Ar mixtures following the initial laser pulse. Points were taken from experimental data, and solid curves show fits of equation \ref{eqn:recomb_rate} to the data to estimate $k_r$, the recombination rate (full details in Materials and Methods).}
  \label{fig:result}
\end{figure}

Laser shock experiments were conducted as described in ref. \citenum{saeidfirozeh2022_jaas} to estimate key parameters of atmospheric ionisation during hypervelocity impact.  Mixtures of Xe-$\rm{H_2}$ and Ar-$\rm{H_2}$ were heated by short pulses of high energy laser to create a plasma.  Ar, as a noble gas with higher ionization threshold than Xe \cite{Zahnle2019_gca}, was chosen so that the relative ionisation efficiencies and recombination rates between the two gases could be compared.  Spectra were recorded from $\sim{200}\,\rm{ns}$ after the initial laser-induced plasma was formed, and from these, the Xe and Ar ionisation fractions were estimated (See Materials and Methods for the complete description; Fig. \ref{fig:result}).  

The laser shock experiments show that Xe is more strongly ionised than Ar (Fig. \ref{fig:result}a vs \ref{fig:result}b): for a plasma temperature of 2$\times10^4\,\rm{K}$, well within the range of temperatures atmosphere is expected to be heated to during impact \cite{silber2018_asr}, $\sim40\%$ of the Xe is present as either singly or doubly charged ions (Fig. \ref{fig:result}a); this compares with less than 1\% of Ar being present as charged ions at the same temperature (Fig. \ref{fig:result}b).  With increasing temperature, more Xe and Ar are ionised, but at all temperatures investigated, Xe remains over an order of magnitude more ionised than Ar.  

The second important result from the experiments is an estimate of the relative  recombination rates of Xe and Ar.  Recombination rates were calculated by fitting the following equation to the data
\begin{equation}
n_e(t) = \dfrac{n_e(0)}{1 + n_e(0) k_r t},
\label{eqn:recomb_rate}
\end{equation}
where $n_e(t)$ ($\rm{cm}^{-3}$) is the electron density at time $t$ (s), $n_e(0)$ is the electron density immediately following the laser pulse (at which time the gas is assumed to be fully ionised), and $k_r$ ($\rm{cm^{3}\,s^{-1}}$) is the recombination rate.  The fits of equation \eqref{eqn:recomb_rate} to these time series are shown in Fig. \ref{fig:result}c and \ref{fig:result}d.  A recombination rate of $6 \pm 2.5 \times 10^{-11} \; {\rm cm^3 \, s^{-1}}$ is found for Xe and a higher rate, $4 \pm 1 \times 10^{-10} \; {\rm cm^3 \, s^{-1}}$, for Ar.  Xe recombination is, therefore found to occur significantly more slowly than Ar recombination.

The results shown in Fig. \ref{fig:result}a and \ref{fig:result}b are for ionization of pure Xe or Ar gases.  However, experiments were also conducted for various mixtures of Xe and Ar with \ce{H2} gas. Each of these species has different ionization energies, $I$ (eV). Since \ce{H2} will have been much more abundant than \ce{Xe} in Mars's early atmosphere \cite{pahlevan2022_epsl,wordsworth2021_ngeo}, and that charge exchange between the ionization products of \ce{H2} ($I = 15.4$ eV) and \ce{Xe} ($I = 12.1$ eV) has been proposed as a mechanism for ionizing \ce{Xe} \cite{Zahnle2019_gca}, we look into the dependency of \ce{Xe} and \ce{Ar} ($I = 15.8$ eV) thermal ionization yield and recombination rates on \ce{H2} concentrations. The results of these experiments (Fig. S2 and S3) show that increasing the $\rm{H_2}$ mixing ratio has only a small effect on the electron density and the recombination rate of Xe and Ar.  This provides reassurance that in an early planetary atmosphere, whether it has a high mean molecular weight e.g., is $\rm{CO_2}$ $\pm$ $\rm{H_2O}$ $\pm$ $\rm{N_2}$ dominated, or is $\rm{H_2}$-rich, Xe's ionization behaviour will be well described by the results for the pure gasses presented in Fig. \ref{fig:result}. 

Overall, the laser shock experiments indicate that atmospheric plasma generated by hypervelocity impact will contain more ionized Xe than Ar and that the Xe will remain ionised for longer than the Ar.  The ionization behaviour of Xe during impact compared to that of Ar therefore favour Xe's transport to high altitude whilst still ionized, and hence its potential loss from the planet along magnetic field lines.  We next take the parameters derived from these experiments and use them in a simple model of impact-driven Xe and Ar escape to investigate whether the observed isotopic fractionation of Xe can be generated and on what timescale (Fig. \ref{fig:abundances}).

\subsection*{Modelling impact-driven Xe loss and fractionation}
\begin{figure}[!ht]
  \includegraphics[width=\textwidth]{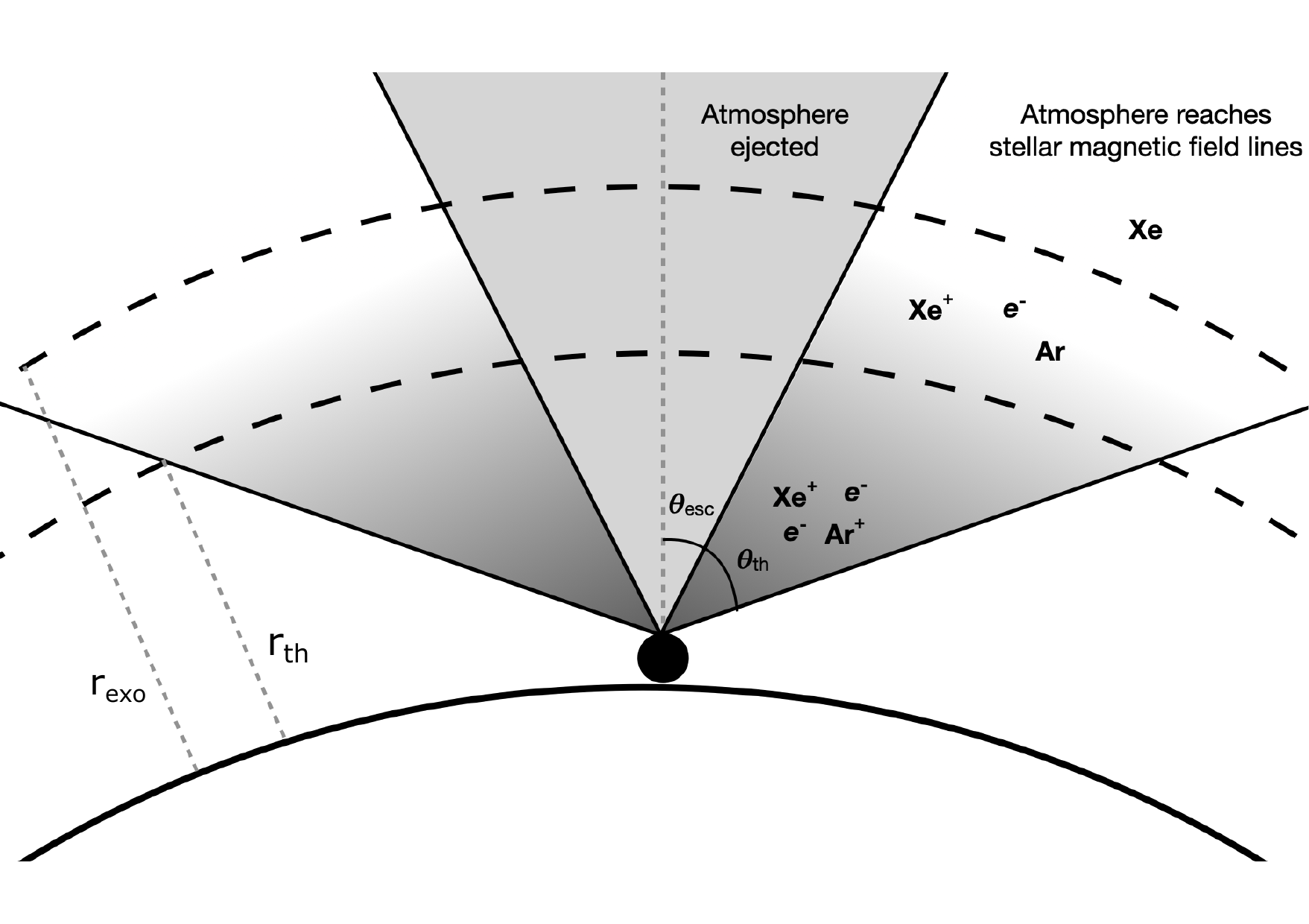}
  \centering
  \caption{{\bf Diagram of xenon impact ejection and escape.} Cross-sections are shown for the two cones of ejected (centre) and ionised and conditionally lost  atmosphere (either side) for our model. Atmosphere in the solid grey cone subtended by $\theta_{\rm esc}$ is ejected from the atmosphere by the impact, is unbound by gravity and moved into the exosphere $> r_{\rm exo}$ (km). Atmosphere in the gradient-grey cone subtended by $\theta_{\rm th}$ passes the homopause and reaches the thermosphere $r_{\rm th}$ (km). Species that remain ionised at this point (e.g., \ce{Xe} and \ce{Ar}) will be confined by the stellar magnetic field and escape the atmosphere.}
  \label{fig:esc-xe-cartoon}
\end{figure}

A schematic of how we model impact-driven loss of Xe to occur is presented in Fig. \ref{fig:esc-xe-cartoon}.  This treatment of impact-driven loss closely follows that developed by ref. \citenum{schlichting2015_icarus}, in which the atmosphere is ejected by planetesimal impact above a tangent plane with respect to the planet's surface.  An atmospheric loss event involves an impactor of a given mass entering the atmosphere with some velocity.  The impactor ejects all atmosphere, i.e., all gases equally, in a solid cone around its entry axis.  A second wider cone extends around this central volume of the lost atmosphere; this outer cone is ionized and ejected upwards, but without sufficient velocity that it will be ballistically ejected from the planet's gravitational well.  Instead, we model gas in this cone of material as being conditionally lost subject to the ionization state of the gases at the point they reach the homopause.  

Before proceeding, we emphasise two important simplifications, or assumptions, of our model's approach.  First, we assume the overall atmospheric pressure during impact bombardment is constant.  In the context of explaining Xe loss and fractionation, this would correspond to a scenario in which impact processing of the martian surface releases major atmospheric constituents such as \ce{CO2} and \ce{H2O}, by melting rock or destabilising ices, equivalent to the mass of atmosphere lost.  Second, we assume that the impactors deliver no noble gases.  This second assumption would correspond to either the impactors themselves containing negligible noble gases when compared to the amount removed (e.g., they may be differentiated and degassed), or, that the gases they do contain are volatilised and ejected during impact in the escaping cone (Fig. \ref{fig:esc-xe-cartoon}) along with the target atmosphere.  We revisit both of these assumptions in the discussion.  

As we saw in the previous section, Xe is ionized more completely and recombines more slowly than Ar.  The key question for impact-driven loss and mass-dependent fractionation of Xe is then to quantify how much more efficiently Xe will be lost compared with the other noble gases (i.e., to explain the observations  in Fig. \ref{fig:abundances}A), and how much more efficiently will the lighter isotopes of Xe be lost than heavier isotopes (i.e., to explain the observations in Fig. \ref{fig:abundances}B).  In the framework of impact-driven escape, these efficiencies are set by the relative degrees of ionization of the noble gases at the point the impact plasma reaches the homopause.  To calculate this our simplified model defines a single pressure in the atmosphere at which the plasma is made.  From this point it calculates the mass of gas in the outer cone that will be ejected upwards but only conditionally lost, and how long it takes that gas to reach the homopause.  Knowing the ionization yields (Fig. \ref{fig:result}A \& B) from impact and the recombination rates (Fig. \ref{fig:result}C \& D), equation \eqref{eq:recomb} can be applied to convert ions back to neutral species during the time it takes to move the impact-induced plasma from its site of formation to the homopause.   

The time to the homopause, therefore, emerges as key in setting the efficiency with which impacts can drive Xe (and Ar) loss.   We assume that the ionized xenon and argon travel a distance $h$ with a time of flight inversely proportional to the square root of their mean molecular weights.  Ions, therefore, are mass segregated in the ionized gas during their time of flight and have more time to become neutralized and then settle back into the lower atmosphere, the heavier they are. We set the time of the flight to
\begin{equation}
t_{\rm fl} = \sqrt{\dfrac{h}{2g}} \, \sqrt{\dfrac{\mu(\ce{X})}{\mu_a}},
\label{eq:tfl}
\end{equation}
where $g$ is the gravitational acceleration, $\mu(\rm{X})$ is the mean molecular weight of species X, and $\mu_a$ is the mean molecular weight of the atmosphere.  As Xe is heavier than argon, equation \eqref{eq:tfl} favours Xe retention and Ar loss.  However, as we have already seen, the recombination rates and initial ionization fractions counteract this effect.  The differential time of flight does, though, mean that the heavier isotopes of Xe may be separated from the lighter isotopes and preferentially retained (Fig. \ref{fig:abundances}B).

Once the partially ionized gas has reached the homopause we assume that all species remaining ionized are lost.  For the purposes of these calculations we are only tracking ionisation-dependent loss of Xe and Ar, but we also track the total atmospheric mass lost by direct ejection (central cone shown in Fig. \ref{fig:esc-xe-cartoon}).  

Our model considers the cone of atmosphere ejected past the homopause into the thermosphere, subtracting out the cone of atmosphere that is entirely ejected (Figure \ref{fig:esc-xe-cartoon}). The fraction of the non-ejected atmosphere that remains \ce{Xe^+} is removed from the atmosphere. This mass loss can be expressed as the depleted mass of Xenon ($\mathcal{M}_D(\ce{Xe})$, kg):
\begin{equation}
\mathcal{M}_D(\ce{Xe}) = \dfrac{\mathcal{M}_0(\ce{Xe})f_0(\ce{Xe})\mathcal{M}_T(\ce{Xe})}{\mathcal{M}_0(\ce{Xe})f_0(\ce{Xe}) + \mathcal{M}_T(\ce{Xe})},
\end{equation}
where $f_0(\ce{Xe})$ is the initial mixing ratio of \ce{Xe} below the homopause, $\mathcal{M}_T(\ce{Xe})$ (kg) is the total mass of \ce{Xe} below the homopause, and:
\begin{equation}
\mathcal{M}_0(\ce{Xe}) = \dfrac{2 N_0}{\beta-2} \, \Bigg(\dfrac{\mu(\ce{Xe})}{\mu_a}\Bigg)\chi_0(\ce{Xe^+})e^{-\Lambda_r}\,\rho_{\rm imp}R_0^3.
\end{equation}
$\chi_0(\ce{Xe^+})$ is the initial fraction of \ce{Xe} that is ionized, $\mu_a$ is the mean molecular weight of the atmosphere, $\mu(\ce{Xe})$ is the mean molecular weight of Xenon, $\rho_{\rm imp}$ (g cm$^{-3}$) is the mass density of the impactor, and $N_0$ and $R_0$ (km) are the number and size of impactors found by integrating the impactor size distribution:
\begin{equation}
dN = \beta N_0 \Bigg(\dfrac{R_0}{r_{\rm imp}}\Bigg)^{\!\!\beta}\,\dfrac{dr_{\rm imp}}{r_{\rm imp}},
\end{equation}
where $r_{\rm imp}$ (km) is the size of a given impactor and $\beta > 2$ is an integer that sets the power-law of the size distribution.

The quantities $R_0$ and $\Lambda_r$ are parameterized in terms of the free and experimentally-constrained parameters of the height of the homopause ($h$, km), the radius of the planet ($R_p$, expressed in Earth radii, $R_{\oplus}$), surface gravity ($g$, m s$^{-2}$), surface pressure ($p$, bar), surface temperature ($T$, K), ionization fraction ($f_e$), and recombination rate ($k_r$, cm$^3$ s$^{-1}$), as so:
\begin{align}
R_0 \, ({\rm km}) =& \; 28.45 \, \big(\mu_a\big)^{1/3}\Bigg(\dfrac{1 \; {\rm g \, cm^{-3}}}{\rho_{\rm imp}}\Bigg)^{\!\!1/3}
\Bigg(\dfrac{h}{1 \, {\rm km}}\Bigg)^{\!\!1/6}
\Bigg(\dfrac{R_{\oplus}}{R_p}\Bigg)^{\!\!1/6}
\Bigg(\dfrac{1 \; {\rm m \, s^{-2}}}{g}\Bigg) \Bigg(\dfrac{p}{1 \, {\rm bar}}\Bigg)^{\!\!1/3}
\Bigg(\dfrac{T}{300 \, {\rm K}}\Bigg)^{\!\!2/3} \\
\Lambda_r =& \; \big(5.4 \times 10^{20} \; {\rm cm^{-3} \, s}\big) \; f_e k_r \Bigg(\dfrac{p}{1 \, {\rm bar}}\Bigg)
\Bigg(\dfrac{300 \, {\rm K}}{T}\Bigg)
\Bigg(\dfrac{h}{1 \, {\rm km}}\Bigg)^{\!\!1/2} \Bigg(\dfrac{1 \; {\rm m \, s^{-2}}}{g}\Bigg)^{\!\!1/2}
\Bigg(\dfrac{\mu(\ce{Xe})}{\mu_a}\Bigg)^{\!\!1/2}.
\end{align}

Details of this model are given in the SI.

\begin{figure}[!ht]
  \includegraphics[width=0.8\textwidth]{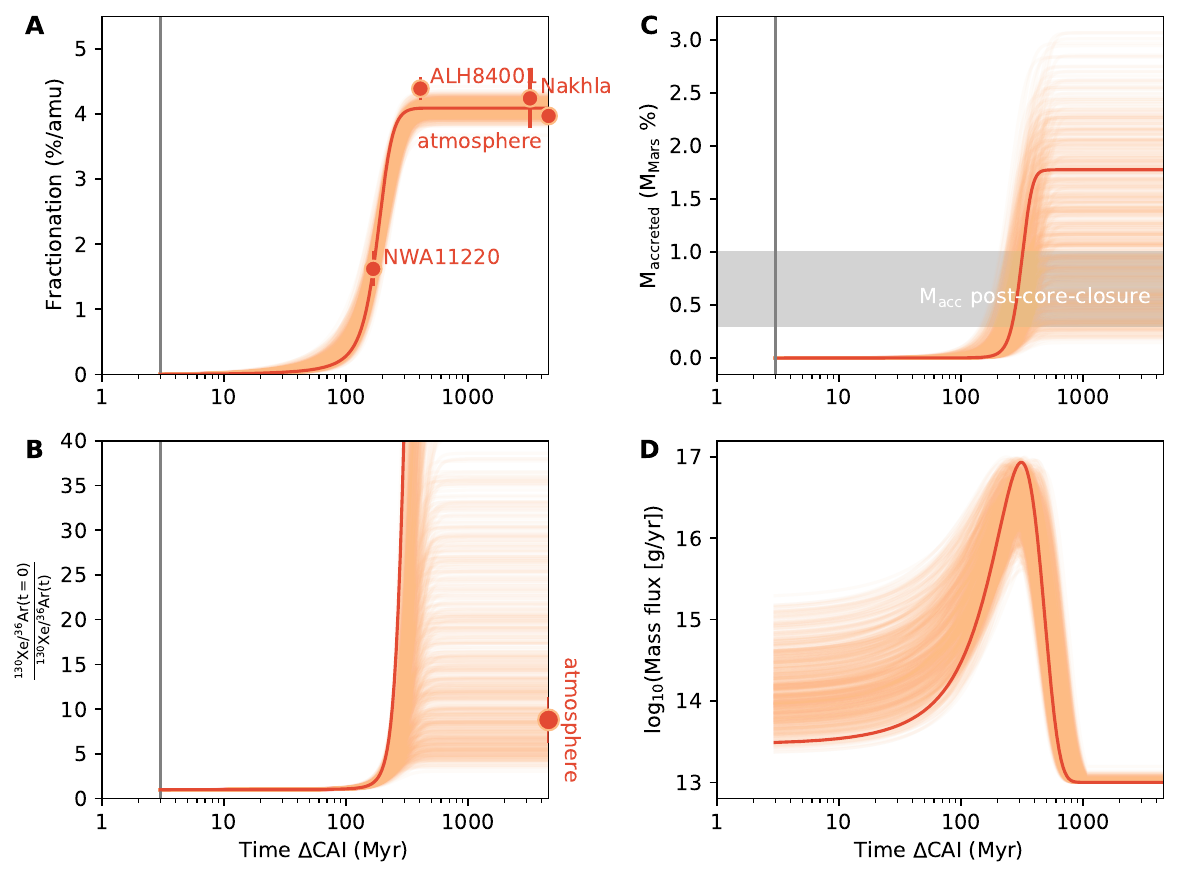}
  \centering
  \caption{{\bf The evolution of martian Xe during impact bombardment.} {\bf (A)}, The isotopic fractionation of Xe through time in the model, compared to modern and ancient estimates of the atmospheric Xe isotope fractionation \cite{gilmour2001_gca,conrad2016_epsl,cassata2017_epsl,cassata2022_epsl}.  {\bf (B)}, The evolution of the Xe/Ar ratio during impact-driven loss.  Impact bombardment in the best fitting model leaves the Xe/Ar ratio below modern values, allowing for subsequent solar-wind driven loss and fractionation of Ar \cite{atreya2013_grl} to drive the atmosphere to modern values.  ({\bf C}), The cumulative mass delivered to Mars in impactors, compared to the estimate of accreted mass post-core closure from HSE's \cite{dale2012_science,day2015_epsl}.  ({\bf D}) The impact rate histories of solutions consistent with the Xe mass fractionation data.  In all cases, the best fitting model (maximum likelihood estimate) is shown in a dark red solid line.}
  \label{fig:composite}
\end{figure}

The forward model described above is combined with a Bayesian Monte Carlo inversion routine to estimate the values for the parameters of interest (see Table \ref{tab:params} for a complete list of parameters and the priors they were assigned in the modelling) by fitting the model to the Xe mass fractionation record (Fig. \ref{fig:composite}A and ref. \citeonline{cassata2022_epsl}).  Bayesian inversion is performed using MultiNest, a Monte Carlo nested sampling algorithm \cite{feroz2009_mnras,feroz2013_arxiv} via pyMultiNest \cite{buchner2014_aa}.  The model includes eight parameters, four of which are included as nuisance parameters to be marginalised over for error propagation.  Four parameters are of key geological interest as they speak to the solar system and martian history, and these are discussed in more detail below.  The value these parameters take when matching the observations is, therefore, a key test of the model's validity; they offer potential for testing against independent geological records.

The first of these parameters is atmospheric surface pressure, $p_{\rm{surf}}$ (bar).  This defines the mass of the martian atmosphere, and therefore it's starting Xe inventory (via the prescribed Xe mixing ratio).  In the context of impact-driven escape, more Xe in the atmosphere initially will require more impacts to remove.  The surface pressure of the martian atmosphere is also of direct relevance to the planet's climate history.  This leads to the second parameter, $C_{\rm{peak}}$, a factor modifying the exponent of the mass flux of impactors onto the planet.  A highly simplified prescription for impactor flux is used, in which a constant background flux is perturbed by a single peak, of height $10^{C_{\rm{}peak}}$ above background (see Materials and Methods for full description).  The timing of this peak in impactor flux is controlled by the third parameter, $t_{\rm{}peak}$, and its width by the fourth parameter $\Delta{}t$.  As we will see, in this simple model the time of the peak impact flux is a key parameter in fitting the temporal evolution of martian Xe loss and fractionation \cite{cassata2022_epsl}

The net effect of the range of $t_{\rm{}peak}$ and $\Delta{}t$ permitted in the inversion (Table \ref{tab:params}) is for there to be a declining impactor flux over the first $\sim1\,\text{Gyr}$ of solar system history.  The integrated impactor flux gives the total mass gained by the planet during impact bombardment, and can in principle be tested against cratering records and highly siderophile element (HSE) derived estimates of post-core formation mass accretion.

\begin{table}[t!]
    \centering
    \caption{\textbf{Parameters estimated in the Bayesian inversion, including their prior values and distributions}. Priors are prescribed as uniform (`U'), log uniform (`LU'), or normally (`N') distributed.}
    \begin{tabular}{ccll} \hline
        Parameter & Units & Description & Prior\\
        \hline
        \multicolumn{4}{l}{Parameters of geological interest}\\
        $p_{\rm{surf}}$ & bar             & surface pressure    & $LU(10^{-3}, 10^1)$\\
        $C_\text{peak}$ & -               & mass flux increase by $10^{C_\text{peak}}$ above background & $U(0, 4)$\\
        $t_\text{peak}$         & Ga              & time of peak mass flux & $U(4.564,4.000)$\\
        $\Delta{}t$ & Gyr               & time window of increased mass flux & $U(0.15, 0.25)$\\
        \multicolumn{4}{l}{Parameters included for error propagation}\\
        $h$           & km              & height of escape    & $U(50, 1000)$\\
        $p$           & bar             & impact pressure in atmosphere & $LU(10^{-6}, p_{\rm{surf})}$\\
        $T$           & K               & temperature of atmosphere between impact and $h$ & $U(200, 1000)$\\
        $f_e$         & -               & ionization fraction of background atmosphere & $LU(10^{-12}, 10^{0})$\\\hline
    \end{tabular}
\label{tab:params}
    \end{table} 

The fit of the model to the time-resolved record of Xe isotopic fractionation data are shown in Fig. \ref{fig:composite}A.  The red line records the solution calculated from the median of the parameters' posterior distributions and provides a close fit to the $\sim4$\,Gyr of Mars's Xe mass fractionation.  In particular, the model is able to reproduce the rapid in-growth of mass-fractionated Xe in Mars's atmosphere in the first few hundred million years of its history.  

The single modern constraint on the factor by which the atmospheric Xe/Ar ratio has been decreased below its initial value, $\left(\frac{\rm{^{130}Xe/^{36}Ar}(t=0)}{\rm{^{130}Xe/^{36}Ar(t)}}\right)$, was not included in the model fit, however solutions range from around the modern atmospheric value to significantly higher degrees of fractionation (Fig. \ref{fig:composite}B; median $\left(\frac{\rm{^{130}Xe/^{36}Ar}(t=0)}{\rm{^{130}Xe/^{36}Ar(t=4.5\,Gyr)}}\right)$ of $\sim100$).  This wide distribution on the level of decrease of the Xe/Ar ratio, yet tightly constrained Xe mass fractionation histories, comes from the sensitivity of the loss of Xe and Ar to their relative recombination rates; a factor that does not affect consideration of Xe isotopes alone, where a given recombination rate can be compensated for through other parameters.  Importantly for the model's consistency with the modern atmosphere, subsequent loss processes may remove atmospheric Ar over Mars's history and thus lower an initially high $\left(\frac{\rm{^{130}Xe/^{36}Ar}(t=0)}{\rm{^{130}Xe/^{36}Ar(t)}}\right)$ down to the observed value \cite{jakosky1994mars}. Given the median $\left(\frac{\rm{^{130}Xe/^{36}Ar}(t=0)}{\rm{^{130}Xe/^{36}Ar(t=4.5\,Gyr)}}\right)$ predicted in our model, a loss of $\sim90\%$ of the post-impact Ar inventory would evolve Mars's atmosphere to its modern Xe/Ar ratio.  This is consistent with independent models explaining the modern $^{36}$Ar/$^{38}$Ar ratio of the martian atmosphere (a ratio affected at less than the part per million level by impact bombardment), which suggest up to 95\% of Mars's atmospheric argon inventory may have been removed gradually over its history \cite{atreya2013_grl}.

\begin{figure}[!ht]
  \includegraphics[width=\textwidth]{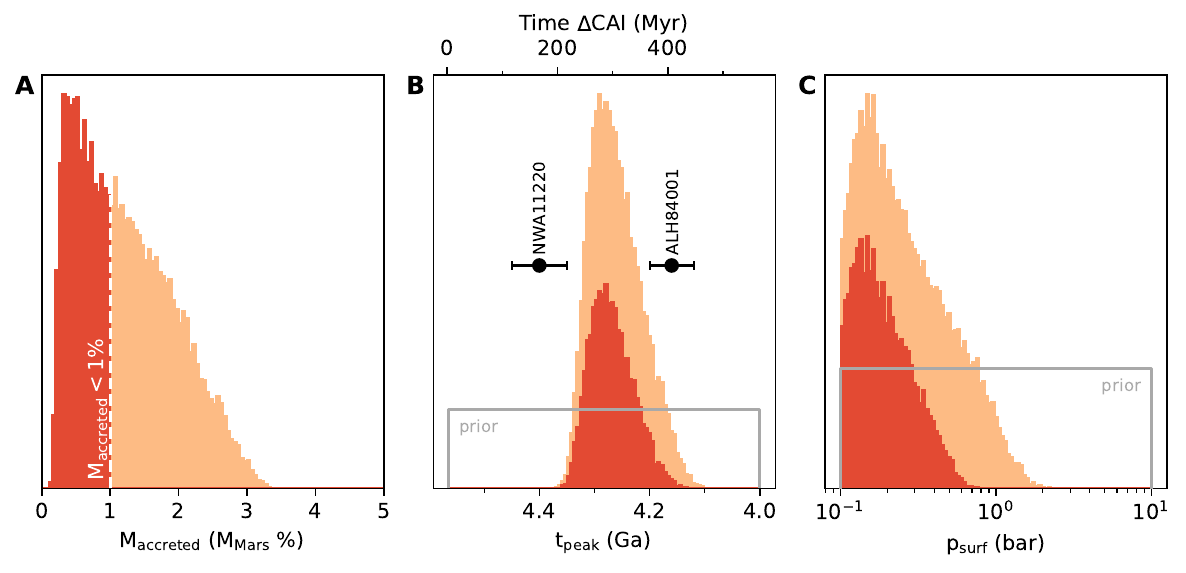}
  \centering
  \caption{{\bf Posterior distributions of parameters consistent with highly siderophile element constraints on accreted mass.} {\bf (A)} The posterior distribution of accreted mass to Mars ($M_{\rm{accreted}}$ ($M_{\rm{Mars}}$ \%)), following impact bombardment sufficient to produce the observed Xe isotopic fractionation.  Approximately 45\% of solutions, given the choice of priors, fall at or below the $\sim1\%$ ($M_{\rm{Mars}}$) limit on post-core closure mass addition inferred from HSE's \cite{dale2012_science,day2015_epsl}, these are shown in light orange.  In dark red are the solutions where less than 1\% the mass of Mars was accreted, and the observed Xe isotopic fractionation was still produced.  {\bf (B)} The posterior distribution of the onset of impact bombardment-driven isotopic fractionation.  The grey line shows the uniform prior, and the histograms show the resultant prior (light orange) and subset of the posterior that matches the accreted mass constraint (dark red).  {\bf (C)} The posterior distribution of surface pressure during the impact bombardment that produces the observed Xe isotopic mass fractionation (light orange).  Dark red identifies the results consistent with constraints on accreted mass.}
  \label{fig:param_favoured}
\end{figure}

\section*{Discussion}
Our experimental and numerical results demonstrate that impact bombardment could have driven the preferential loss and isotopic fractionation of Xe in the atmosphere of Mars.  Critically, this could only have occurred early in the lifetime of the solar system, whilst the mass flux of impactors onto planets was still significant: the ability of the models to match the observed rapid in-growth of Xe isotope fractionation followed by stasis (Fig. \ref{fig:composite}C), directly emerges from this history.  Specifically, the models require an early peak in impactor flux followed by rapid decline in impact rate over the first few hundred million years of solar system history \cite{marchi2014_nature}.

Two important assumptions the model makes are of a constant atmospheric pressure during bombardment and no Xe delivery. Atmospheric bombardment is expected to erode atmospheres \cite{schlichting2015_icarus}, inferred from the same theoretical approach we take to predict Xenon escape. This erosion can be counter-balanced by delivery of volatiles \cite{Sinclair2020}, or by degassing from the magma generated at the impact site. For simplicity, we assume that these different sources (delivery, impact-generated magma degassing) and sinks (impact erosion), when summed over all species lead to a stable, albeit low pressure, atmosphere, and that the sources provide negligible Xenon (the convergent state of an atmosphere under bombarded \cite{Sinclair2020}). Below we consider the effect of relaxing each of these assumptions.

Allowing changes in the atmospheric surface pressure can enhance impact-driven isotope fractionation. If the atmosphere is being eroded, then the Xenon depletion will be increased because, as we have shown (Fig. \ref{fig:param_favoured}C), Xenon ionization and escape is more efficient when the surface atmospheric pressure is lower. If the atmosphere is growing, due to efficient delivery or efficient magma degassing, then our mechanism will be frustrated. Further work to constrain these sources and sinks, and explore potentially observable implications of these constraints, will be needed to inform whether and how this assumption should be relaxed.  However, the evidence for a low pressure atmosphere on early Mars suggests significant atmospheric growth during this bombardment epoch was unlikely (see discussion below).

We consider the assumption of no Xe delivery during bombardment a reasonable starting point for modelling the following reasons.
\begin{enumerate}
\item Impactor Xe will be lost along with the cone of atmosphere it is interacting with. The efficiency of impactor Xe partitioning between loss and delivery is difficult to predict with our present state of knowledge, because it will depend on the dynamics of the impact event. If the delivered Xenon is included within the cone of matter that is ejected from the atmosphere \cite{schlichting2015_icarus}, then it will necessarily not contribute to the atmospheric budget. A complete physical model would be required to calculate the partitioning of impactor Xe between loss and delivery, which is beyond the scope of this work, but qualitatively has the effect of diminishing the potential of Xe delivery compared with the impactor's initial inventory;

\item For a surface atmospheric pressure of $0.1\,\text{bar}$, Xenon will only be added if the delivery brings more Xenon than $10^{-11}\,$g/g (i.e., more Xe contained in the meteorite than is lost above the homopause; \cite{Scherer2000}.  If we consider the higher surface pressures permitted by the model (Fig. \ref{fig:param_favoured}C), then even higher impactor Xe concentrations are required to perturb the atmospheric Xe budget. By comparison to a conservative $10^{-11}\,$g/g value though, carbonaceous chondrites bracket this, having a range of Xenon concentrations between $(0.01 - 2.5) \times 10^{-11}\,$g Xe/g total. If the more Xenon-rich Carbonaceous chondrites represent the average impactor composition during the tail end of accretion, then Xenon will be slowly increased by impact delivery.  Ordinary chondrites, in contrast, have less than $10^{-11}\,$g/g Xenon on average \cite{Schelhaas1990}, and so will not contribute enough Xenon to substantially change our results. E-Chondrites have even less Xenon, on average $5 \times 10^{-13}\,$g/g \cite{Crabb1981}.  Evidence points to Mars's late accretion being dominated by ordinary- and enstatite chondrite-like material \cite{dauphas2024_icarus}, suggesting Xe delivery would have been unimportant even before consideration of point (1) above; and,

\item Mars’s atmosphere is constrained to have started with a solar-like Xe isotopic composition \cite{pepin2000_ssr}, from which subsequent atmospheric evolution occurred.  This observation rules out late delivery as a prominent contributor to Mars’s atmospheric Xe, because this would have delivered non-solar chondritic Xe --- impacts must have acted primarily to sculpt Mars's initial solar-Xe inventory, rather than add to it.
\end{enumerate}

In addition to impactor delivery, impact-generated magma degassing is unlikely to provide significant Xenon to the atmosphere: Xenon's insolubility in magmas \cite{carroll1994_revmin} makes it likely that the Xenon concentration in the Mars's crust/interior will be much lower than the Xenon abundance in the atmosphere.  Overall then, we consider it most likely that Mars's Xe inventory during this bombardment epoch was dominated by loss and fractionation, not delivery or interior outgassing. 

An important aspect of impact bombardment as a means of preferentially driving Xe loss and fractionation is that it does not depend on background $\rm{H_2}$ in a planet's atmosphere.  This is in contrast to previous work for Earth and Mars \cite{Zahnle2019_gca,cassata2022_epsl}, in which Xe ionisation by charge exchange and transport through the atmosphere is linked to atmospheric hydrogen content: such that in the case of the Earth the cessation of Xe fractionation is hypothesised to be linked to the oxidation of the atmosphere \cite{Zahnle2019_gca}.  Our experimental results showed limited sensitivity of Xe ionisation and recombination to the presence of background $\rm{H_2}$ and the impact mechanism directly provides ballistic transport of Xe through the atmosphere to the ionosphere.  Hence, impact-driven fractionation and Xe loss does not make specific predictions, nor have specific requirements for, the evolution of the background atmospheric composition.  A corollary of this is that end of Xe fractionation is also separated from atmospheric $\rm{H_2}$ evolution, instead presumably being linked to declining bombardment flux or a waning of the martian magnetic field.  The latter scenario cannot explain when Xe fractionation ceases, because ALH84001, which records a Xe isotopic composition close to the modern martian atmosphere also preserves a magnetic field \cite{weiss2002_epsl,cassata2022_epsl}.  However, this serves to emphasise that for both the model we propose, and those previously considered \cite{cassata2022_epsl}, the presence of a martian magnetic field is critical in guiding ionised Xe from the planet whilst leaving non-ionised gases behind.  Future insights into the history of the martian dynamo will therefore provide tests of these models of Xe escape.

A key geological prediction impact-driven Xe fractionation makes is of the mass of late delivery to Mars|.  Estimates from HSE abundances in Mars's mantle, inferred from martian meteorites, place the mass of material delivered post-core closure to be $\sim0.4-1\%$ of the mass of the planet \cite{dale2012_science,day2015_epsl}.  Fig. \ref{fig:composite}C calculates the integrated mass delivered over time for the median solution (red line) and a sample of the posteriors (light orange).  Many of these solutions fall above the window of permissible mass delivery; however, a significant fraction ($\sim45\%$) have parameters governing the loss and fractionation of Xe that allow for smaller impactor mass fluxes.  Impact-driven loss is, therefore, in principle, able to operate within the independent constraints on mass delivery to Mars to explain the history of its atmospheric Xe mass fractionation.

We can look more closely at the properties of those simulations that successfully match the mass-delivery constraint to understand the broader requirements for conditions in the early solar system and on Mars if impact bombardment is to explain the Xe fractionation.  Fig. \ref{fig:param_favoured} highlights the posterior distributions for the mass accreted to Mars over the early period of intense bombardment ($M_{\rm{}accreted}$), the timing of the impact peak ($t_\text{peak}$), and the atmospheric surface pressure during the bombardment interval ($p_\text{surf}$).  For Models accreting $<1\%$ the mass of Mars ($\sim45\%$ of simulations), in line with the HSE observations, the posteriors are shown in dark orange.  The key insight from these posterior distributions is that a narrow timing of the onset of Xe fractionation and lower atmospheric pressure on Mars ($<1\,\text{bar}$) is favoured for successful model runs.  The reasons for these parameter values being favoured are that (1) too early an onset of atmospheric escape leads to too much Xe isotopic fractionation occurring before the $\sim4.4\,\text{Ga}$ NWA11220-derived constraint on martian atmospheric Xe \cite{cassata2022_epsl}, and (2) increased atmospheric pressure suppresses the degree of Xe fractionation observed, as a smaller fraction of the initial Xe pool ends up processed by impacts.  Combined, these aspects of the impact-drive loss scenario favour Mars's atmosphere beginning to record fractionation from $\sim200\,Myr$ after the birth of the solar system, with a low pressure atmosphere at this time.

Early peaks in impactor rate have long been predicted by dynamical models seeking to match the architecture of the solar system \cite{tsiganis2005_nature,gomes2005_nature}.  However, there has been much debate over the timing of such events.  Measurements of Xe isotopes in martian meteorites NWA11220 and ALH84001, dated at $\sim4.4$ and $4.16\,\text{Ga}$ respectively, tightly constrain the timing of impact-driven fractionation.  Between these two measurements, mass fractionation of Xe isotopes increases from $\sim1.6$ to $\sim4.4\,\%/\text{amu}$, essentially the modern value of the Mars atmosphere \cite{conrad2016_epsl,cassata2017_epsl, cassata2022_epsl}.  In our simplified model this epoch of Xe fractionation must be matched directly by a peak of impact bombardment.  However, more complex models that couple a time evolving atmospheric mass with monotonically declining impactor flux (an `accretion tail' scenario \cite{neukum2001_chron}) could likely also be reconciled to the data: such models have been previously suggested for Mars's atmospheric evolution \cite{melosh1989_nature,scherf2021_ssr}.  In light of our impact-based description of the Xe isotope data, accretion-tail scenarios would need to be paired with an atmospheric mass stabilising at $\sim4.35\,\text{Ga}$ to an initial solar composition from which isotopic fractionation could in-grow.

Even if in principle the model cannot separate an accretion tail scenario from am impact peak, our results do place an upper limit on how long after solar system birth any giant planet instability and associated bombardment flux can occur.  Xe fractionation, and therefore intense impact sculpting of the atmosphere, needs to end by 300--$400\,\text{Myr}$ after CAI formation, else martian Xe would experience a more protracted period of isotopic evolution than it exhibits.  This is an important bound on the timing of giant planet instability, given the difficulty of directly probing impact history this far back in time \cite{norman2009_elements,walton2023_gca}.

An important feature of impact-driven Xe loss and its relation to Mars's impact chronology is that it is most efficient for small impactors, those just above the threshold to eject atmosphere \cite{schlichting2015_icarus}.  Larger impactors, such as those responsible for large scale basin formation (e.g., Hellas basin) are both less efficient at removing atmosphere per unit mass, and eject \emph{all} atmosphere above the tangent plane to the planet \cite{schlichting2015_icarus}, thereby being less efficient at fractionating the atmosphere.  Whereas it is the \emph{differential} loss of atmosphere that is possible with smaller impactors that drives Xe fractionation.  In this sense, Mars's history of Xe fractionation, if driven by impacts, is a complementary archive of impact bombardment to the surface geological record: whereas the latter best preserves and age dates \cite{fassett2011_icarus,bottke2017_ngeo} the large events, the former integrates the effects of the much more numerous small events. How these large basin forming impactors could have affected atmospheric Xe is through remelting of the crust and mantle, processes which may release trapped Xe.  In fact, a late (i.e., young) age of these large basin forming events may help explain a peculiar feature of the martian Xe isotopic record: that in our quantification of the Xe fractionation (Fig. \ref{fig:composite}A), the martian meteorites record a peak fractionation at ALH84001 followed by a decline between then and the modern atmosphere (and, albeit with lower certainty, Nahkla).  Large impacts liberating Xe from the martian mantle, or delivering some solar-like Xe, could explain this slight reversal in fractionation trend.

An impact solution to Mars's fractionated Xe also has implications for its climate.  Our results favour a low pressure atmosphere, $<1\,\text{bar}$, remaining after formation and impact bombardment (Fig. \ref{fig:param_favoured}C).  This result is consistent with independent constraints on Mars's early atmospheric pressure
\cite{kite2014_ngeo,scherf2021_ssr}, and the observation of sulfur mass-independent isotopic fractionation in martian meteorite NWA11220 \cite{endo2019_grl,tomkins2020_gca}, which favours low atmospheric pressures.  Importantly, a tenuous atmosphere is consistent with a climate history of punctuated warmth and episodes of surface liquid water, with the later tail of impacts contributing to this stochasticity \cite{wordsworth2021_ngeo}.

Impact-driven loss and fractionation of Xe may also be tested by NASA's planned DAVINCI mission to Venus \cite{glaze2017davinci}, which has as a core science aim measurement of the heavy noble gases in the planet's atmosphere.  If modern Venus is representative of the planet's past, then its Xe isotopes should not show preferential fractionation compared to Kr, and the Kr to Xe ratio should be sub-chondritic rather than super-chondritic as on Mars.  This is because Venus's massive atmosphere would stifle the efficiency of impact-driven fractionation of Xe: First, we have shown that for a thick atmosphere too little Xe is removed to perturb the existing inventory; and second, both the model presented here and that of ref. \citenum{Zahnle2019_gca} require a magnetic field to channel ions away from the planet, the thick atmosphere and resulting hot surface and slow mantle cooling may have suppressed Venus's dynamo for its entire history preventing this process.  Conversely, in this paradigm, if DAVINCI's measurements do evidence preferential Xe loss and fractionation, then it may point to a more temperate early Venus (e.g., as propsed by ref. \citenum{way2016venus}). 

The history of Xe on Mars has importance for Earth's own Xe depletion.  Our experiments and calculations show the potential of impactors to fractionate Xe early in a planet's history.  This impact-driven fractionation occurs too early to explain the slow ingrowth of Xe mass fractionation seen in Earth's atmosphere \cite{avice2018_gca}.  However, recent results have shown the presence of a more ancient history of Xe fractionation in Earth's deep mantle \cite{peron2022_epsl}.  This signal may point to Earth having once experienced similar impact loss processes to Mars, albeit resolvable only in mantle long hidden from subsequent volatile addition and recycling.

\newpage\clearpage

\section*{Materials and Methods}

\subsection*{Apparatus}

The plasma UV–ViS measurements were performed inside a vacuum-sealed cylindrical glass cell. The sample was contained within the cell using three diagnostic quartz windows and an in-house SwagelokTM gas–vacuum handling system. The Nd:YAG 1064 \,nm laser, capable of delivering a maximum energy of 850 mJ, was employed to generate a plasma spark at the vessel's center. This laser radiation was focused using a coated plano-convex quartz lens (= 1.5 cm, f = 10.5 cm). The emission spectra of the laser-induced plasma were then captured by the ESA 4000 Echelle spectrograph (LLA Instruments GmbH, Germany) using a fiber optic cable. A photodiode was attached to detect the laser spark radiation.

The laser pulses were cumulated at a repetition frequency of 10 Hz to simulate an impact shock wave. The observation gate delay was adjusted to varying values to facilitate time-resolved screening while maintaining a constant gate width of 500 ns. The time gating was controlled using ESAWIN software (version 14.3.0).

A vacuum pump and pressure gauge from Pfeiffer Vacuum Austria GmbH were employed to manage the gas flow and measure pressure within the cell. A schematic of the experimental setup can be found in the Supplementary Materials (Fig. S1).

\subsection*{Experimental Conditions }
The vessel was filled with pure noble gases and their mixtures under a series of nominal pressures and concentrations investigated in this study: pressures of 40 - 700\,Torr and hydrogen concentration ranges from 0 up to 90 \%. For any measurements, certified gas samples, i.e., respectively 5.6 Linde Gas argon, 5.0 Linde Gas xenon, and 6.0 Linde Gas di-hydrogen were used. The spectra have been recorded by an Echelle spectrograph with delays of 10, 500, 1000, 1500, 2000, 2500, and 3000\,ns with the gate width of 100\,ns. 

\subsection*{Experimental data acquisition and processing}
Approximately 250 spectra were recorded in the wavelength range between 200 to 750\,nm. A simple pre-processing procedure \cite{saeidfirozeh2022_jaas} was applied to all spectra to obtain basic plasma diagnostics.  This analysis was performed by in-house programmed scripts in \textsc{python-numpy}, and \textsc{python-scipy}. Theoretical values for the calculations were extracted from the NIST database \cite{kramida2014nist}.

\section*{Acknowledgements}
William Cassata is thanked for his guidance in interpreting the data reporting the noble gas isotope composition of the martian atmosphere.  Guillaume Avice is thanked for the discussion of the terrestrial Xe record.  This work is part of a research series funded by grant no. 21-11366S of the Czech Science Foundation

\subsection*{Funding} 
This work is part of a research series funded by grant no. 21-11366S of the Czech Science Foundation. 

\section*{Author Contributions} 
All authors discussed the experimental plan.  HS MF designed experiments and conducted them with VL PK and LP.  All authors contributed to the drafting and editing of the manuscript.

\section*{Competing interests} All authors declare that they have no competing interests.

\section*{Data availability} All data needed to evaluate the conclusions in the paper are present in the paper and the Supplementary Materials.


\newpage\clearpage
\end{document}